 \documentclass[12pt]{article}
 \usepackage {graphicx}
 \usepackage[latin1]{inputenc}
 \usepackage{amsmath}
 \usepackage{amsfonts}
 \usepackage{amssymb}
 \usepackage{amsthm}
 \date{}
 \newcommand{\be}{\begin{equation}}
 \newcommand{\ee}{\end{equation}}
 \newcommand{\ba}{\begin{eqnarray}}
 \newcommand{\ea}{\end{eqnarray}}
 \newcommand{\inc}{{\it{i}}}

 \newcommand{\efbold}{\mbox{{\boldmath $\vec f$}}}
 
 \newcommand{\erbold}{\mbox{{\boldmath $\vec r$}}}
 
 \newcommand{\omegabold}{\mbox{{\boldmath $\vec \omega$}}}

  \newcommand{\vbold}{\mbox{{\boldmath $\vec v$}}}
  \newcommand{\Vbold}{\mbox{{\boldmath $\vec V$}}}

 \newcommand{\fbold}{\mbox{\boldmath $\vec {\boldmath{\,f}}$}}

\begin{document}
 \title{
 {\Large{\textbf{The Theory of Bodily Tides.\\
 The Models and the Physics}
            }}}
 \author{
 {\Large{Michael Efroimsky}}\\
 {\small{US Naval Observatory, Washington DC 20392 USA}}\\
 {\small{e-mail: michael.efroimsky$\,$@$\,$usno.navy.mil~}}\\
 }
 \maketitle
 \begin{abstract}
 \noindent
 Description of tides is based on the form of dependence of the geometric lag $\delta_1$ on the tidal-flexure frequency $\chi$.
 Some authors put $\delta_1\sim\chi^0$, others set $\delta_1\sim\chi^1$. The actual dependence of the lag on the frequency is
 complicated and determined by the planet's rheology. A particular form of this dependence will fix the form of the frequency
 dependence of the tidal quality factor $\,Q\,$. Since at present we know the shape of the function $Q(\chi)$, we can reverse our
 line of reasoning and obtain the appropriate frequency-dependence of the lag. Employment of a realistic dependency $Q(\chi)$
 renders considerable changes in the timescales defined by tidal dynamics.

 \end{abstract}

 \section{The Singer-Mignard theory of bodily tides}

  The earliest efforts aimed at modeling bodily tides were undertaken in the end of the XIX$^{th}$ century  by Darwin (1879, 1880). A
  much simplified theory was later offered by MacDonald (1964) who assumed the geometric lag angle to be a fixed
  constant. With the latter assumption regarded as a critical deficiency, his theory soon fell into disuse in favour of the approach
  by Singer and Mignard. Singer (1968) suggested that the subtended angle should be proportional to the principal frequency $\chi$ of
  the tide. This was equivalent to setting $\Delta t$ constant in (\ref{2} - \ref{3}). Singer applied this new theory to the Moon and to
  Phobos and Deimos. A detailed mathematical development of Singer's idea can be found in Mignard (1979, 1980) who completely avoided
  using the lag angle and operated only with the position and time lags. Later, Singer's assumption of a constant $\,\Delta t\,$ was
  employed also by Touma \& Wisdom (1994) and Peale \& Lee (2000).

 A far more comprehensive theory was developed by Kaula (1964). While Kaula assumed the lag angle and the quality factor to be
 frequency-independent constants, his approach was general enough to embrace an arbitrary frequency-dependence of these parameters.
 This way, it enables one to employ an arbitrary rheology, i.e., to assume that different tidal modes fall behind the appropriate
 modes of the perturbing potential with different phases and, accordingly, with difference geometric lag angles.

 In the special case of all tidal modes lagging by the same time delay $\,\Delta t\,$, one can model a dynamical tide with a static
 tide where all the time-dependent variables are shifted back by $\,\Delta t\,$. This gives birth to the concept of a {\it{fictitious
 satellite}}.

 If the actual satellite is located at a planetocentric position $\erbold$, it generates a
 tidal bulge that either advances or retards the satellite motion, depending on the
 interrelation between the planetary spin rate $\omega_p$ and the tangential part of
 satellite's velocity $\vbold$ divided by $r\equiv|\erbold|$. It is convenient to
 imagine that the bulge emerges beneath a fictitious satellite located at
 \ba
 \erbold_f\,=\,\erbold\,+\,\fbold\;\;\;,
 \label{1}
 \label{401}
 \ea
 where the position lag $\,\fbold\,$ is given by
 \ba
 {\vec{\mbox{\it\textbf{f}}}}\;=\;\Delta t\;\left(\;\omegabold_p\times
 \erbold\;-\;\vbold\;\right)\;\;\;,
 \label{2}
 \label{402}
 \ea
 $\Delta t\,$ being the time lag between the real and fictitious
 tide-generating satellites. For vanishing eccentricity and inclination, the quantity
 \ba
 \delta\;\equiv\;\frac{|\fbold |}{r}\;=\;\frac{\Delta t}{r}\;
 |\omegabold_p\times\erbold\;-\;\vbold|\;\;\;.
 \label{3}
 \label{403}
 \ea
 \begin{center}
 \begin{figure}
 \includegraphics[width=7.8in]{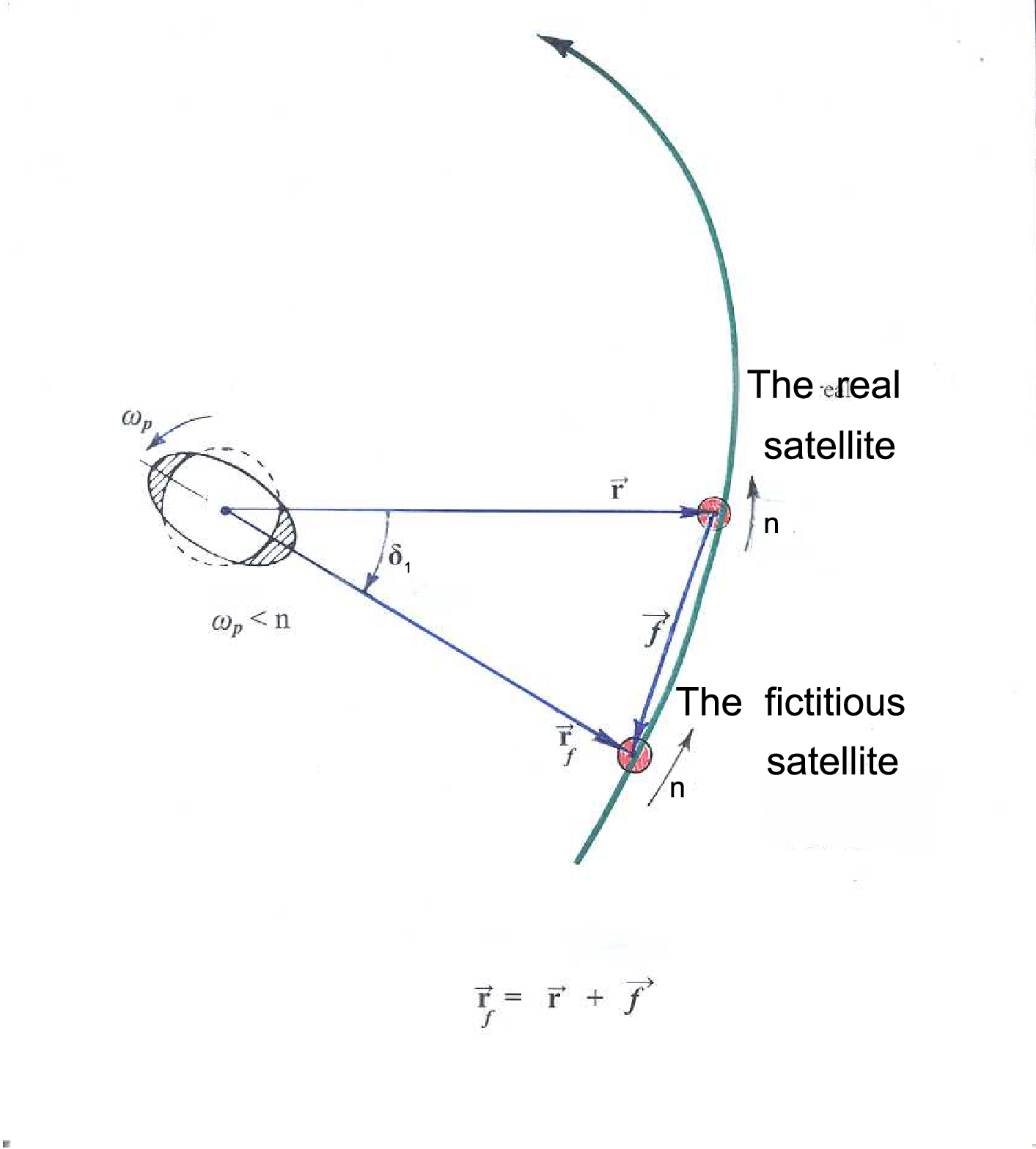}
 \caption{The position lag $\,\efbold\,$ and the angular lag $\,\delta_1\,$
 of a satellite located below the synchronous orbit. (For a moon
 located above the synchronous orbit both lags will be pointing in an opposite
 direction.) Be mindful that $\,\delta\,$ defined through (\ref{3}) does not
 coincide with the subtended angle $\,\delta_1\,$, except for circular orbits.}
 \label{Tidal_picture}
 \end{figure}
 \end{center}
 coincides with the absolute value of the angle
 $\,\delta_1\,$ subtended at the planet's centre between the satellite and the
 tidal bulge, as on Fig.~\ref{Tidal_picture}. For non-circular orbits, they differ.

 The imaginary satellite is merely a way of illustrating the time lag
 between the tide-raising potential and the distortion of the body. This
 concept implies no new physics, and is but a
 convenient figure of speech employed to express the fact that at each
 instance of time the dynamical tide is modeled with a static tide where
 all the time-dependent variables are shifted back by $\,\Delta t\,$, i.e.,
 (a) the moon is rotated back by $\,\vbold\,\Delta t\,$, and (b) the
 attitude of the planet is rotated back by $\,\omegabold_p\,\Delta t\,$.
 From the viewpoint of a planet-based observer, this means that a dynamical
 response to a satellite located at $\,\erbold\,$ is modelled with a static
 response to a satellite located at $\,\erbold_f\,\equiv\,\erbold\,-\,
 \Delta t\,(\vbold\,-\,\omegabold_p\times\erbold)\,$.

 It should be reiterated that the simplistic concept of a fictitious satellite can be used only for an equatorial circular orbit, within
 the Singer-Mignard model (the latter implying that all tidal modes get delayed by the same time lag $\,\Delta t\,$).

 Delayed response means lagging both in the location of the bulge and in its height (compared to the height of a static bulge).  The
 well-known relation $\;1/Q\,=\,\tan 2\delta_1\;$ interconnects the quality factor with the ``tangential lag" $\,\delta_1\,$ (see the
 Appendix), while the magnitude lagging is taken care of  by the frequency-dependence of the Love number.

 \section{The tidal frequency $\chi$ and the quality factor $Q$}

 \subsection{The principal tidal frequency}

 The principal frequency of tidal flexure is $\,2\,|n-\omega_p|\,$, where $\,n\,$ and $\omega_p\equiv|\,\omegabold_p|$ stand for the
 satellite's mean motion and the planet's spin rate.

 For an equatorial circular orbit, the satellite velocity relative to the surface is
 \ba
 |\,\omegabold_p\times\erbold~-~\vbold\,|~=\;r\;|\,\omega_p\,-\,n\,|\;\;\;,
 \label{4}
 \label{404}
 \ea
 the tidal frequency being
 \ba
 \chi\;=\;2\;|\,\omega_p\,-\,n\,| \;\;\;\;,
 \label{5}
 \label{405}
 \ea
 and the angular lag being
 \ba
 \delta\;\equiv\;\frac{\Delta t}{r}\;|\,\omegabold_p\times\erbold\;-\;
 \vbold\,|\;=\;\frac{\Delta t}{2}\;\chi\;\;\;.
 \label{6}
 \label{406}
 \ea
 While in this particular case we have only the principal tidal mode, in the generic case (when $\,\inc\,$ or $\,e\,$ is not small, or
 when $\,\Delta t\,$ is frequency-dependent) an infinite spectrum of tidal modes will emerge.

 \subsection{Goldreich's admonition:\\ a general difficulty stemming from nonlinearity}

 Darwin (1908), Jeffreys (1961), Kaula (1964) and other authors always started with the fact that each elementary volume of the planet
 is subject to a tide-raising potential, which in general is not periodic but can be expanded into a sum of periodic terms. They then
 employed the linear approximation introduced by Love, according to which the tidal perturbations of the potential yield linear response
 of the shape and linear variations of the stress. Within the linear approximation, the overall dissipation inside the planet may be
 represented as a sum of attenuation rates corresponding to each periodic disturbance:
 \ba
 \langle\,\dot{E}\;\rangle\;=\;\sum_{i}\;\langle\,
 \dot{E}(\chi_{\textstyle{_i}})\;\rangle
 \label{407}
 \ea
 where, at each frequency $\,\chi_i\,$,
 \ba
 \langle\,\dot{E}(\chi_{\textstyle{_i}})~\rangle~=~-~2~\chi_{\textstyle{_i}}~
 \frac{\,\langle\,E(\chi_{\textstyle{_i}})~\rangle\,}{Q(\chi_{\textstyle{_i}})
 }~=\,\;-\;\chi_{\textstyle{_i}}\;\frac{\,E_{_{peak}}(\chi_{\textstyle{_i}})
 \,}{Q(\chi_{\textstyle{_i}})}\;\;\;,
 \label{408}
 \ea
 $\langle\,.\,.\,.\,\rangle~$ standing for averaging over flexure cycle,
 $\,E(\chi_{\textstyle{_i}})\,$ denoting the energy of deformation at
 the frequency $\,\chi_{\textstyle{_i}}\,$, and $Q(\chi_{\textstyle{_i}})\,$
 being the quality factor of the material at this frequency. Introduced
 empirically as a means to figleaf our lack of knowledge of the attenuation
 process in its full complexity, the notion of $\,Q\,$ has proven to be practical
 due to its smooth and universal dependence upon the frequency and temperature.
 At the same time, this empirical treatment has its predicaments and
 limitations. Its major inborn defect was brought to light by Peter
 Goldreich who pointed out that the attenuation rate at a particular
 frequency depends not only upon the appropriate Fourier component of the
 stress, but also upon the {\emph{overall}} stress. This happens because
 for real minerals each quality factor $\,Q(\chi_{\textstyle{_i}})\,$
 bears dependence not only on the frequency $\,\chi_{\textstyle{_i}}\,$,
 but also on the $\,\chi_{\textstyle{_i}}\,$ component of the stress and,
 most importantly, also on the {\emph{overall}} stress. This, often-neglected,
 manifestation of nonlinearity may be tolerated only when the
 amplitudes of different harmonics of stress are comparable. However, when
 the amplitude of the principal mode is orders of magnitude higher than
 that of the harmonics (tides being the case), then the principal mode
 will, through this nonlinearity, make questionable our entire ability to
 decompose the overall attenuation into a sum over frequencies. Stated
 differently, the quality factors corresponding to the weak harmonics will
 no longer be well defined physical parameters.

 Here follows a quotation from Goldreich (1963):\\
 {\emph{``... Darwin and Jeffreys both wrote the tide-raising potential
 as the sum of periodic potentials. They then proceeded to consider the
 response of the planet to each of the potentials separately. At first
 glance this might seem proper since the tidal strains are very small
 and should add linearly. The stumbling block in this procedure, however,
is the amplitude dependence of the specific dissipation function. In
the case of the Earth, it has been shown by direct measurement that
$\,Q\,$ varies by an order of magnitude if we compare the tide of
frequency  $\,2\omega - 2n \,$ with the tides of frequencies $\,
2\omega - n \,$,  $\, 2\omega - 3n \,$, and $\,\frac{\textstyle
3}{\textstyle 2}n\,$. This is because these latter tides have
amplitudes which are smaller than the principle tide (of frequency
$\, 2\omega - 2n \,$ ) by a factor of eccentricity or about 0.05. It
may still appear that we can allow for this amplitude dependence of
Q merely by adopting an amplitude dependence for the phase lags of
the different tides. Unfortunately, this is really not sufficient
since a tide of small amplitude will have a phase lag which
increases when its peak is reinforcing the peak of the tide of the
major amplitude. This non-linear behaviour cannot be treated in
detail since very little is known about the response of the planets
to tidal forces, except for the Earth."}}

 On these grounds, Goldreich concluded the paragraph with an important
 warning that we {\emph{``use the language of linear tidal theory, but we
 must keep in mind that our numbers are really only parametric fits to a
 non-linear problem."}}

 In order to mark the line beyond which this caveat cannot be ignored, let
 us first of all recall that the linear approximation remains applicable
 insofar as the strains do not approach the nonlinearity threshold, which
 for most minerals is of order $\,10^{-6}\,$. On approach to that
 threshold, the quality factors may become dependent upon the strain
 magnitude. In other words, in an attempt to extend the expansion
 (\ref{407} - \ref{408}) to the nonlinear case, we shall have to introduce,
 instead of $\,Q(\chi_{\textstyle{_i}})\,$, some new functions
 $\,Q(\,\chi_{\textstyle{_i}}\,,\;E_{_{peak}}(\chi_{\textstyle{_i}})\,,
 \;E_{_{overall}}\,)\;$. (Another complication is that in the nonlinear
 regime new frequencies will be generated, but we shall not go there.)
 Now consider a superposition of two forcing stresses -- one at
 the frequency $\,\chi_{\textstyle{_1}}\,$ and another at
 $\,\chi_{\textstyle{_2}}\,$.
 Let the amplitude $\;E_{_{peak}}(\chi_{\textstyle{_1}})\;$ be close or
 above the nonlinearity threshold, and
 $\;E_{_{peak}}(\chi_{\textstyle{_2}})\;$ be by an order or two
 of magnitude smaller than $\;E_{_{peak}}(\chi_{\textstyle{_1}})\;$.
 To adapt the linear machinery (\ref{407} - \ref{408}) to the
 nonlinear situation, we have to write it as
 \ba
 \langle\,\dot{E}~\rangle~=~\langle\,\dot{E}_1~\rangle~+~\langle\,\dot{E}_2
 ~\rangle~=~-~\chi_{\textstyle{_1}}~\frac{~E_{_{peak}}(\chi_{\textstyle{_1}
 })~}{Q(\,\chi_{\textstyle{_1}}\,,~E_{_{peak}}(\chi_{\textstyle{_1}})\,)}~
 -~\chi_{\textstyle{_2}}~\frac{\,E_{_{peak}}(\chi_{\textstyle{_2}})\,}{Q(
 \chi_{\textstyle{_2}}\,,\;E_{_{peak}}(\chi_{\textstyle{_1}})\,,\;
 E_{_{peak}}(\chi_{\textstyle{_2}})\,)}\;\;\;,\;\;\;\;\;
 \label{409}
 \ea
 the second quality factor bearing a dependence not only upon the frequency
 $\,\chi{\textstyle{_2}}\,$ and the appropriate magnitude $\;E_{_{peak}}(
 \chi_{\textstyle{_2}})\;$, but also upon the magnitude of the
 {\emph{first}} mode, $\;E_{_{peak}}(\chi_{\textstyle{_1}})\;$, -- this
 happens because it is the first mode which makes a leading contribution
 into the overall stress. Even if (\ref{409}) can be validated as an
 extension of (\ref{407} - \ref{408}) to nonlinear regimes, we should
 remember that the second term in (\ref{409}) is much smaller than the
 first one (because we agreed that $\;E_{_{peak}}(\chi_{\textstyle{_2}})\,
 \ll\,E_{_{peak}}(\chi_{\textstyle{_1}})\;$). This results in two quandaries.
 The first one (not mentioned by Goldreich) is that a nonlinearity-caused
 non-smooth behaviour of $\;Q(\,\chi_{\textstyle{_1}}\,,\;E_{_{peak}}(
 \chi_{\textstyle{_1}})\,)\;$ will cause variations of the first term in
 (\ref{409}), which may exceed or be comparable to the entire second term.
 The second one (mentioned in the afore quoted passage from Goldreich) is
 the phenomenon of nonlinear superposition, i.e., the fact that the
 smaller-amplitude tidal harmonic has a higher dissipation rate (and,
 therefore, a larger phase lag) whenever the peak of this harmonic is
 reinforcing the peak of the principal mode. Under all these circumstances,
 fitting experimental data to (\ref{409}) will become a risky
 business. Specifically, it will become impossible to reliably measure the
 frequency dependence of the second quality factor; therefore the entire
 notion of the quality factor will, in regard to the second frequency, become
 badly defined. This difficulty limits our ability to extend (\ref{407} - \ref{408}) to nonlinear
 regimes.

 The admonition by Goldreich had been ignored until an alarm sounded. This happened when
 the JPL Lunar-ranging team applied the linear approach to determining the
 frequency-dependence of the Lunar quality factor, $\,Q(\chi)\,$. They obtained (or, as
 Peter Goldreich rightly said, {\emph{fitted}} their data to) the dependency $\,Q(\chi)\,
 \sim\,\chi^{\alpha}\,$ (Williams et al. 2001). The value of the exponential for the Moon
 turned out to be {\emph{negative}}: $\,\alpha\,=\,-\,0.07\;$, a result firmly tabooed by
 the condensed-matter physics for this range of frequencies (Karato 2008).

 One possible approach to explaining this result may be the following. Let us begin with
 a very crude estimate for the tidal strains. For the Moon, the tidal displacements are
 of order $\,0.1m\,$ (zero to peak).
 The most rough estimate for the
 strain can be obtained through dividing the displacement by the radius of the body. While
 this ratio is almost twenty times less than the
 $\,10^{-6}\,$ nonlinearity threshold, we should keep in mind that in reality the
 distribution of the tidal strain is a steep function of the radius, with the strain
 getting its maximum near the centre of the body, as can be seen from equations (48.17)
 in Sokolnikoff (1956). The values of strain can vary in magnitude, over the radius, by
 about a factor of five.\footnote{~Accordingly, the tidal-energy radial distribution,
 too, is strongly inhomogeneous, with a maximum in the planet's centre -- see Fig. 3 in
 Peale \& Cassen (1978).} Another order of magnitude, at least, will come from the fact
 that the Moon is inhomogeneous and that its warmer layers are far more elastic than its
 rigid surface. As a result, deep in the Lunar interior the tidal strain will exceed the
 afore mentioned nonlinearity threshold. Hence the uncertainties in determination of
 $\,Q(\chi)\,$ by the Lunar-ranging team.

 An alternative approach to explanation of the unusual frequency-dependence of the lunar quality factor would come from the observation
 that the tidal $\,Q\,$ is not the same as the seismic $\,Q\,$. While the latter should scale, at all quantities, as a positive power
 of the forcing frequency, the former must change its behaviour at the lowest frequencies and must diverge in the zero-frequency limit.
 The latter is obvious physically, because the tidal torque should not diverge on crossing of the synchronous orbit.

 In what follows, we shall adhere to the linear approximation. It will also be implied that the frequencies are not too small, so we do
 not have to care about the difference between the seismic and tidal quality factors. This difference is to be addressed elsewhere.

 \section{Dissipation in the mantle.}

 \subsection{Generalities}

 Back in the 60s and 70s of the past century, when the science of low-frequency seismological measurements was yet under development,
 it was widely thought that at long time scales the quality factor of the mantle is proportional to the inverse frequency. This fallacy
 proliferated into planetary astronomy where it was received most warmly, because the law $\,Q\,\sim\,1/\chi\,$ turned out to be the
 only model for which the linear decomposition of tides gives a set of bulges displaced from the direction to the satellite by the
 same angle. Any other frequency dependence $\,Q(\chi)\,$ entails superposition of bulges corresponding to the separate frequencies,
 each bulge being displaced by its own angle. This is the reason why the scaling law $\,Q\,\sim\,1/\chi\,$, long disproved and abandoned
 in geophysics (at least, for the frequency band of our concern), still remains a pet model in celestial mechanics of the Solar system.

 Over the past twenty years, a considerable progress has been achieved in the low-frequency seismological measurements, both in the lab
 and in the field. Due to an impressive collective effort undertaken by several teams, it is now a firmly established fact that {\it{for
 frequencies down to a certain threshold (about $\sim 1$ yr$^{-1}$, for the Earth's mantle) the quality factor of the mantle is
 proportional to the frequency to the power of a {\textbf{positive}} fraction}} $\,\alpha\,$. This dependence holds for all rocks within
 a remarkably broad band of frequencies: from several MHz down to about $\;1\;$yr$^{-1}\,$.

 At timescales longer than $\,1$ yr, all the way to the Maxwell time (about $\,100
 $~yr), attenuation in the mantle is defined by viscosity, so that the
 quality factor is, for all minerals, well approximated with $\,{\eta}\chi/M\,$,
 where $\eta$ and $M$ are the sheer viscosity and the sheer elastic modulus of the
 mineral. Although the values of both the viscosity coefficients and elastic moduli
 greatly vary for different minerals and are sensitive to the temperature, the
 overall quality factor of the mantle still scales linear in frequency.

 The linear law $\,Q\,\sim\,\chi\,$ extends all the way down to the zero-frequency limit, insofar as the Maxwell or near-Maxwell
 behaviour extends thereto. We should keep in mind, though, that in the said limit the {\it{tidal}} quality factor will differ
 from its seismic counterpart, due to self-gravitation of the body.\\
 ~\\
 All in all, we have:\\
 \ba
 \mbox{For}\;\;\;\;\;\;10^7\;\;\mbox{Hz}\;\;>\,\;\chi\;\;>\;\;1\;yr^{-1}\;:\;\;\;Q\,\sim\,
 \chi^{\alpha}\;,\,\;\mbox{with}\;\,\alpha\,=\,0.2\,-\,0.4\;\;\;(\,0.2\;
 \mbox{for partial melts})\;.\;\;\;\,
 \label{one}
 \label{422}
 \ea
 \ba
 \mbox{For}\;\;\;\;\;~~\,1\,\;yr^{-1}\,>\,\chi\,>\,10^{-2}\;yr^{-1}\;:\;\;\;\;\;Q\,
 \sim\,\chi\;\;\;.\;\;\;\;~~~~~~~~~~~~~~~~~~~~~~~~~~~~~~~~~~~~\,~~~~~~~~~~~~~
 ~~~~~~~~~
 \label{two}
 \label{423}
 \ea
  \ba
 \mbox{For}\;\;\;\;\;10^{-2}\;yr^{-1}\,>\,\chi\;:\;\;\;\mbox{the~$\,${\it{seismic}}$\,$~(not~{\it{tidal}}!)~quality~factor~still~scales~as~~}
 Q\,\sim\,\chi\;.\;\;\;\;\;~~~~~~~~
 \label{three}
 \label{424}
 \ea
 It is important to emphasise that the positive-power scaling law (\ref{one}) is well proven not only for samples in the lab but also
 for vast seismological basins and, therefore, is universal. Hence, this law may be extended to the tidal friction, up to the caveat
 mentioned in (\ref{424}).

 Generally, in the study of tidal dynamics, the complex case of low frequencies is encountered only in the problems related to
 entrapment into or crossing of spin-orbit resonances. It the current paper, we shall not approach this topic, and thus shall be
 interested solely in the frequency range described in (\ref{one})

 Below we provide an extremely squeezed review of the published data whence the scaling
 law (\ref{one}) was derived by the geophysicists. The list of sources will be incomplete,
 but a full picture can be restored through the further references contained in the works
 to be quoted below. For a detailed review on the topic, see Chapter 11 of the book by
 Karato (2008) that contains a systematic introduction into the theory of and experiments
 on attenuation in the mantle.

 \subsection{Circumstantial evidence: attenuation in minerals.\\
  Laboratory measurements and some theory}

 Even before the subtleties of solid-state mechanics with or without melt are brought up,
 the positive sign of the power $\,\alpha\,$ in the dependence $\,Q\,\sim\,\chi^{\alpha}
 \,$ may be anticipated on qualitative physical grounds. For a damped oscillator obeying
 $\;\ddot{z}\,+\,2\,\beta\,\dot{z}\,+\,\chi^2\,z\,=\,0\;\,$, the quality factor is equal
 to $\,\chi/(2\beta)\,$, i.e., $\,Q\,\sim\,\chi\,$.

 Solid-state phenomena causing attenuation in the mantle may be divided into
 three groups: the point-defect mechanisms, the dislocation mechanisms, and
 the grain-boundary ones.

 Among the point-defect mechanisms, most important is the transient diffusional
 creep, i.e., plastic flow of vacancies, and therefore of atoms, from one grain
 boundary to another. The flow is called into being by the fact that vacancies
 (as well as the other point defects) have different energies at grain boundaries
 of different orientation relative to the applied sheer stress. This anelasticity
 mechanism is wont to obey the power law $\,Q\,\sim\,\chi^{\alpha}\,$ with
 $\,\alpha\,\approx\,0.5\,$.

 Anelasticity caused by dislocation mechanisms is governed by the viscosity law
 $\,Q\,\sim\,\chi\,$ valid for sufficiently low frequencies (or sufficiently high
 temperatures), i.e., when the viscous motion of dislocations is not restrained by
 the elastic restoring stress.\footnote{~At higher frequencies or/and lower
 temperatures, the restoring force ``pins" the defects. This leads to the
 law $\,Q\,\sim\,(\textstyle{1\,+\,\tau^2\chi^2})\tau^{-1}\chi^{-1}\,$,
 parameter $\,\tau\,$ being the relaxation time whose values considerably
 vary among different mechanisms belonging to this group. As the mantle is
 warm and viscous, we may ignore this caveat.}

 The grain-boundary mechanisms, too, are governed by the law $\,Q\sim\chi^{\alpha}$,
 though with a lower exponent: $\,\alpha\approx 0.2 -0.3$. This behaviour gradually
 changes to the viscous mode ($\alpha=1$) at higher temperatures and/or at lower
 frequencies, i.e., when the elastic restoring stress reduces.

 We see that in all cases the quality factor of minerals should grow with frequency.
 Accordingly, laboratory measurements confirm that, within the geophysically
 interesting band of $\,\chi\,$, the quality factor behaves as $\,Q\,\sim\,
 \chi^{\alpha}\,$ with $\,\alpha\,=\,0.2\,-\,0.4\;$. Such measurements have been
 described in Karato \& Spetzler (1990) and Karato (1998). Similar results were
 reported in the works by the team of I. Jackson -- see, for example, the paper
 (Tan et al 1997) where numerous earlier publications by that group are also
 mentioned.

 To this we would add that in aggregates with partial melt the frequency
 dependence of $\,Q\,$ keeps the same form, with $\,\alpha\,$ leaning to
 $\,0.2\,$ -- see, for example, Fontaine et al (2005)
 and references therein.

 \subsection{Direct evidence: attenuation in the mantle.\\
 Measurements on seismological basins}

 Low-frequency measurements, performed by different teams over various basins of the Earth's
 upper mantle, agree on the pivotal fact: the seismological quality factor scales
 as the frequency to the power of a {\emph{positive}} fraction $\,\alpha\;$ -- see,
 for example, Mitchell (1995), Stachnik et al (2004), Shito et al (2004), and
 further references given in these sources.\footnote{~So far, Figure 11 in Flanagan
 \& Wiens (1998) is the only experimental account we know of, which only partially
 complies with the other teams' results. The figure contains two plots depicting
 the frequency dependencies of $\,1/Q_{sheer}\,$ and $\,1/Q_{compress}\,$. While
 the behaviour of both parameters remains conventional down to $\,10^{-1}\,$Hz,
 the sheer attenuation surprisingly goes down when the frequency decreases to
 $\,10^{-3}\,$Hz. Later, one of the Authors wrote to us that ``{\emph{Both P and
 S wave attenuation becomes greater at low frequencies. The trend towards lower
 attenuation at the lowest frequencies in Fig. 11 is not well substantiated.}}"
 (D. Wiens, private communication) Hence, the consensus on (\ref{one}) stays.}

 \subsection{Tidal dissipation versus seismic dissipation}

 One of the basic premises of our model will be an assumption that, for terrestrial
 planets, the frequency-dependence of the $\,Q\,$ factor of bodily tides is similar
 to the frequency-dependence (\ref{one} - \ref{two}) of the seismological $\,Q\,$
 factor. This premise is based on the fact that the tidal attenuation in the mantle
 is taking place, much like the seismic attenuation, mainly due to the mantle's
 rigidity. This is a nontrivial statement because, in distinction from earthquakes,
 the damping of tides is taking place both due to rigidity and self-gravity of the
 planet. Modelling the planet with a homogeneous sphere of density $\,\rho\,$,
 rigidity $\,\mu\,$, surface gravity $\,\mbox{g}\,$, and radius $\,R\,$, Goldreich
 (1963) managed to separate the rigidity-caused and self-gravity-caused inputs into
 the overall tidal attenuation. His expression for the tidal quality factor has the
 form
 \ba
 Q\;=\;Q_o\left(\;1\;+\;\frac{2}{19}\;\frac{\mbox{g}\;\rho\;R}{\mu}\;\right)\;\;\;,
 \label{Goldreich}
 \ea
 $Q_o\,$ being the value that the quality factor would assume were self-gravity absent
 (i.e., were damping due to rigidity only). To get an idea of how significant the
 self-gravity-produced input could be, let us plug there the mass and radius of Mars
 and the rigidity of the Martian mantle. For the Earth mantle, $\,\mu\,=\,65\,\div\,
 80\,$GPa. Judging by the absence of volcanic activity over the past hundred(s)
 of millions of years of Mars' history, the temperature of the Martian upper mantle is
 (to say the least) not higher than that of the terrestrial one. Therefore we may safely
 approximate the Martian $\,\mu\,$ with the upper limit for the rigidity of the
 terrestrial mantle: $\,\mu\,=\,8\,\times\,10^{10}\,$Pa. All in all, the relative
 contribution from self-gravity will look as
 \ba
 \frac{2}{19}\;\frac{\mbox{g}\;\rho\;R}{\mu}\;=\;
 \frac{6}{76\,\pi}\;\frac{\gamma\;M^2}{\mu\;R^4}\;
 \approx\;\frac{1}{40}\;\frac{(\;6.7\,\times\,10^{\textstyle{^{-11}}}\,
 \mbox{m}^{\textstyle{^3}}\,\mbox{kg}^{\textstyle{^{-1}}}\,\mbox{s}^{\textstyle{^{-2}}}\;)
 \;\;(\;6.4\,\times\,10^{\textstyle{^{23}}}\,\mbox{kg}\;)^{\textstyle{^2}}}{(\;
 11^{\textstyle{^{10}}}\;\mbox{Pa}\;)\;\;(3.4\,\times\,10^{\textstyle{^6}}\,m)^{
 \textstyle{^4}}}\;\approx\;5.2\,\times\,10^{\textstyle{^{-2}}}\;\;,\;\;\;\;
 \label{estimate}
 \ea
 where $\gamma\,$ stands for the gravity constant. This rough estimate shows that
 self-gravitation contributes, at most, several percent into the overall count of
 energy losses due to tides. This is the reason why we extend to the tidal $\,Q\,$
 the frequency-dependence law measured for the seismic quality factor.

  The above estimate is conservative, because we approximated the Martian $\,\mu\,$
  with the terrestrial value of $\,\mu\,$. If the interior of Mars is cooler than
  that of the Earth, then its rigidity should be higher, and therefore
  the ratio (\ref{estimate}) may reduce to less than one percent.

 \subsection{The frequency and the temperature}

 In the beginning of the preceding subsection we already mentioned that though the
 tidal $Q$ differs from the seismic one, both depend upon the frequency in the same
 way, because this dependence is determined by the same physical mechanisms. This
 pertains also to the temperature dependence, which for some fundamental reason
 combines into one function with the frequency dependence.

 As explained, from the basic physical principles, by Karato (2008, 1998), the
 frequency and temperature dependencies of $\,Q\,$ are inseparably connected.
 Since the quality factor is dimensionless, it must retain this property despite
 the exponential frequency dependence. This may be achieved only if $\,Q\,$ is a
 function not of the frequency {\emph{per se}} but of a dimensionless product of
 the frequency by the typical time of defect displacement. This time exponentially
 depends upon the activation energy $\,A^*\,$, whence the resulting function reads
 as
 \ba
 Q\;\sim\;\left[\,\chi\;\exp(A^*/RT)\,\right]^{\alpha}\;\;\;.
 \label{427}
 \ea
 For most minerals of the upper mantle, $A^*$ lies within the limits of
 $360-540$ kJ mol$^{-1}$. For example, for dry olivine it is about $520$
 kJ mol$^{-1}$.

 Thus through formula (\ref{427}) the cooling rate of
 the planet plays a role in the orbital evolution of satellites: the lower
 the temperature, the higher the quality factor and, thereby, the smaller
 the lag $\, \delta\,$. For the sake of a crude estimate, assume
 that most of the tidal attenuation is taking place in some layer, for which
 an average temperature $T$ and an average activation energy $A^*$ may be
 introduced.
 Then from (\ref{427}) we have: $\,\Delta Q/{Q}\,\approx\,\alpha\,A^*/RT\,$.
 For a reasonable choice of values $\,\alpha=0.3\,$ and $\,A^*=\,5.4\,\times
 \,10^5\,J/mol\,$, a drop of the temperature from $\,T_o=2000\,K\,$ down by
 $\,\Delta T=200\,K\,$ will result in $\,\Delta Q/Q\,\approx\,1\,$. So a
 $\,10\,\%\,$ decrease of the temperature can result in an about $\,100\,\%\,$
 growth of the quality factor.

 \section{Conclusions}

 While in the MacDonald theory the geometric lag is postulated to be constant, in the Singer-Mignard theory of tides it is
 the time lag that is assumed constant. However, neither of these two choices conform to the geophysical data within the frequency band
 of our interest. As the angular lag $\,\delta_1\,$ is inversely proportional to the tidal $\,Q\,$ factor, the frequency-dependence of
 both $\,\delta_1\,$ and $\,\Delta t\,$ is unambiguously defined by the frequency-dependence of $\,Q\,$.

 According to numerous studies, the quality factor obeys the law $\,Q\,\sim\,\chi^{\alpha}\,$, where $\,\alpha\,$ lies within
 $\,0.2\,-\,0.4\,$. For this reason, the time lag $\,\Delta t\,$, employed in the Singer-Mignard theory as a fixed parameter, in
 reality is a function of the frequency. Through that, $\,\Delta t\,$ becomes a function of the current values of the orbital elements
 of the satellite, and so becomes $\delta_1$. This clause alters the integration of the planetary equations.

 A numerical simulation of the Mars-Phobos dynamics, in neglect of the primary's cooling and in neglect of the tides on the satellite,
 will be presented in Efroimsky \& Lainey (2007). Those computations show that Phobos will fall on Mars in 45 Myr from now. This
 estimate is 50 percent longer than the one following from the tidal models used in the past. The difference originates from our using
 the realistic frequency dependence of attenuation. This demonstrates that the currently accepted time scales of dynamical evolution
 should be reexamined using the actual frequency dependence of the lags.

  ~\\

 {{\textbf{\Large{Acknowledgments}}}}\\
 ~\\
 I wish to deeply thank Valery Lainey, George Kaplan, Victor Slabinski, and
 S. Fred Singer for our fruitful discussions on the theory of tides. My
 very special gratitude goes to Shun-ichiro Karato who so kindly consulted
 me on the theory and phenomenology of the quality factor.

 ~\\

 ~\\

{{\textbf{\Large{Appendix.~~~The quality factor and the angular lag}}}}\\
 ~\\
  ~\\
 In this Appendix, we shall recall why the quality factor $\,Q\,$ is inversely proportional to the phase lag, and why the phase lag is
 twice the geometric lag angle.\\
 ~\\
 ~\\
 \noindent
 {\textbf{{A.1~~The case of a circular equatorial orbit.}}}\\
  ~\\
 We shall begin with the simple case of an equatorial moon on a circular orbit. At
 each point of the planet, the tidal potential produced by this moon will read
  \ba
  W\;=\;W_o\;\cos \chi t\;\;\;,
  \label{A3}
  \label{469}
  \ea
 the instantaneous frequency being given by
 \ba
 \chi\,=\,2~|n\;-\;\omega_p|~~~.~~~
 \label{A3}
 \label{470}
 \ea
 Let $\,\mbox{g}\,$ denote the free-fall acceleration. An element of the planet's volume
 lying beneath the satellite's trajectory will then experience a vertical elevation of
 \ba
 \zeta\;=\;\frac{W_o}{\mbox{g}}\;\cos (\chi t\;-\;2\delta_{_1})\;\;\;.
 \label{A4}
 \label{471}
 \ea
 Accordingly, the vertical velocity of this element of the planet's volume will amount to
  \ba
 u\;=\;\dot{\zeta}\;=\;-\;\chi\;\frac{W_o}{\mbox{g}}\;\sin (\chi t
 \;-\;2\;\delta_{_1})\;=\;-\;\chi\;\frac{W_o}{\mbox{g}}\;\left(\sin \chi
 t\;\cos 2\delta_{_1}\;-\;\cos \chi t\; \sin 2\delta_{_1}\right)\;\;.\;\;
 \label{A5}
 \label{472}
 \ea
 The satellite generates two bulges -- on the facing and opposite sides of the planet -- so each
 point of the surface is uplifted twice through a cycle. This entails the factor of two in the
 expressions (\ref{470}) for the frequency. The phase in (\ref{A4}), too, is doubled,
 though the necessity of this is less evident.\footnote{~Let $\,x\,$ signify a position along
 the equatorial circumference of the planet. In the absence of lag, the radial elevation at a
 point $\,x\,$ would be:
 \ba
 \nonumber
 \zeta\;=\;\frac{W_o}{\mbox{g}}\;\cos
 k(x\;-\;v\;t)\;\;\;,\;\;\;\;\;\;v\,=\,R\,\sigma\;\;\;,
 \ea
 $v\,$ being the velocity of the satellite's projection on the ground,
 $\,R\,$ being the planet's radius, and
 $\sigma$ being simply $\,|n-\omega_p|\,$ because we are dealing with a
 circular equatorial orbit. The value of $\,k\,$ must satisfy
 \ba
 \nonumber
 k\;v\;=\;2\;\sigma\;\;\;,\;\mbox{i.e.,}\;\;\;\;k\;v\;=\;\chi\;\;\;,
 \ea
 to make sure that at each $\,x\,$ the ground elevates twice per
 an orbital cycle. The above two formulae yield:
 \ba
 \nonumber
 k\;R\;=\;2\;\;\;.
 \ea
 In the presence of lag, all above stays in force, except that the
 formula for radial elevation will read:
 \ba
 \nonumber
 \zeta\;=\;\frac{W_o}{\mbox{g}}\;\cos k(x\;-\;v\;t\;+\;D)\;\;\;,\;\;\;\;
 \mbox{where}\;\;\;\; D\;=\;R\;\delta_{_1}\;\;\;,
 \ea
 $D\,$ being the linear lag, and $\,\delta_{_1}\,$ being the angular
 one. Since $\,k\,v\,=\,2\,$, we get:
 \ba
 \nonumber
 \cos \left[\,k\;(x\;-\;v\;t\;+\;R\;\delta_{_1})\,\right]\;=\;\cos \left[\,k\;x
 \;-\;k\;v\;t\;+\;k\;R\;\delta\,\right]\;=\;\cos
 \left[\,k\;x\;-\;(k\;v\;t\;-\;2\;\delta_{_1})\,\right]\;\;\;,
 \ea
 so that, at some fixed point (say, at $\,x\,=\,0\,$) the elevation becomes:
 \ba
 \nonumber
 \zeta(t)\;=\;\frac{W_o}{\mbox{g}}\;\cos (k\;v\;t\;-\;2\;\delta_{_1})\;\;\;.
 \ea
 We see that, while the geometric lag is $\;\delta_{_1}\,$, the phase lag is
 double thereof.}

 The energy dissipated over a time cycle $\,T\,=\,2\pi/(\chi)\,$, per
 unit mass, will, in neglect of horizontal displacements,\footnote{~In Appendix
 A.2, it will be explained why the horizontal displacements may be neglected.} be
 \ba
 \nonumber
 \Delta E_{_{cycle}} &=& \int^{T}_{0}u\left(-\,\frac{\partial W}{
 \partial r}\right)dt=
 \,-\left(-\,\chi \frac{W_o}{\mbox{g}}\right)\,\frac{\partial W_o}{
 \partial r}\int^{t=T}_{t=0}\cos \chi t\,\left(\sin \chi t\,
 \cos 2\delta_{_1}\,-\,\cos \chi t\, \sin 2\delta_{_1}\right)dt\\
 \nonumber\\
 \nonumber\\
 &=&\,-\;\chi\;\frac{W_o}{\mbox{g}}\;\frac{\partial W_o}{\partial r}\;\sin 2\delta_{_1}\;
 \frac{1}{\chi}\;\int^{\chi t\,=\,2\pi}_{\chi t\,=\,0}\;\cos^2 \chi t\;\;d(\chi t)\;=\;-\;
 \frac{W_o}{\mbox{g}}\;\frac{\partial W_o}{\partial r}\;\pi\;\sin 2\delta_{_1}\;\;,\;\;\;
 ~~~~~~~~~~~~~~~~~~~~
 \label{A6}
 \label{}
 \ea
 while the peak energy stored in the system during the cycle will read:
 \ba
 \nonumber
 E_{_{peak}}&=&\int^{T/4}_{0} u \left(-\,\frac{\partial W}{
 \partial r}\right)dt =
 \,-\left(-\,\chi\,\frac{W_o}{\mbox{g}}\right)\frac{\partial W_o
 }{\partial r}\int^{t=T/4}_{t=0}\cos \chi t\,\left(\sin
 \chi t\,\cos 2\delta_{_1}\,-\,\cos \chi t\,\sin 2\delta_{_1}\right)dt\\
 \nonumber\\
 \nonumber\\
 \nonumber
 &=&\;2\;\sigma\;\frac{W_o}{\mbox{g}}\;\frac{\partial W_o}{\partial r}\;\left[\;
 \frac{\cos 2\delta_{_1}}{\chi}\;\int^{\chi t\,=\,\pi/2}_{\chi t\,=\,0}
 \;\cos \chi t\;\sin \chi t\;\;d(\chi t)\;-\;\frac{\sin 2\delta_{_1}
 }{\chi}\;\int^{\chi t\,=\,\pi/2}_{\chi t\,=\,0}\;\cos^2 \chi t
 \;\;d(\chi t)\;\right]\\
 \nonumber\\
 \nonumber\\
 &=&\;\frac{W_o}{\mbox{g}}\;\frac{\partial W_o}{\partial r}\;\left[\;\frac{1}{2}
 \;\cos 2\delta_{_1}\;-\;\frac{\pi}{4}\;\sin 2\delta_{_1}\;\right]~~~~~~~~~~~~~~~~~~~
 ~~~~~~~~~~~~~~~~~~~~~~~~~~~~~~~~~~~~~~~~~~~~~~~~~~~~~~~
 \label{A7}
 \label{}
 \ea\\
 whence
 \ba
 Q^{-1}\;=\;\frac{-\;\Delta E_{_{cycle}}}{2\,\pi\,E_{_{peak}}}\;=\;\frac{1}{2\,\pi}\;\,
 \frac{\pi\;\sin 2\delta_{_1}}{~\frac{\textstyle 1}{\textstyle 2}\;\cos 2\delta_{_1}\;-\;
 \frac{\textstyle\pi}{\textstyle 4}\;\sin 2 \delta_{_1}}\;\approx\;\tan 2 \delta_{_1}\;\;\;.
 \label{A8}
 \label{}
 \ea
 ~\\
 ~\\
 {\textbf{{A.2~~Final touch}}}\\

 In our derivation of the interrelation between $\,Q\,$ and $\,\delta_1\,$, we greatly simplified the situation, taking into account
 only the vertical displacement of the planetary surface, in response to the moon's pull. Here we shall demonstrate that this
 approximation is legitimate, at least in the case when the planet is modeled with an incompressible homogeneous medium.

 As a starting point, recall that the tidal attenuation rate within the planet is
 well approximated with the work performed on it by the satellite:\footnote{~A small
 share of this work is continuously being spent for decelerating the Earth rotation.}

 The power $\,P\,$ exerted by a tide-raising secondary on its primary can be written as
 \ba
 P\;=\;-\;\int \,\rho\;\Vbold\;\cdot\;\nabla W\;d^3x
 \label{A9}
 \ea
 $\rho\,,\;\Vbold\,$, and $\,W\,$ signifying the density, velocity, and tidal potential in the small volume $~d^3x~$ of the primary.
 The mass-conservation law $~\nabla\cdot(\rho\Vbold)\,+\frac{\textstyle \partial \rho}{\textstyle\partial t}\,=\,0\,~$ enables one to
 shape the dot-product into the form of
 \ba
 \rho\,\Vbold\cdot\nabla W\,=\,
 \nabla\cdot(\rho\,
 \Vbold\,W)\,-\,\rho\,W\,\nabla\cdot\Vbold\,-\,\Vbold\,W\,\nabla\rho\;\;~.\;\;\;\;
 \label{}
 \ea
 Under the realistic assumption of the primary's incompressibility, the term with $\,\nabla\cdot\Vbold\,$ may be omitted. To get rid
 of the term with $\,\nabla \rho\,$, one has to accept a much stronger approximation of the primary being homogeneous.
 Then the power will be rendered by
 \ba
 P\;=\;-\;\int\,\nabla\,\cdot\,(\rho\;\Vbold\;W)\,d^3x
 \;=\;-\;\int\,\rho\;W\;\Vbold\,\cdot\,{\vec{\bf{n}}}\;\,dS\;\;\;,
 \label{A9}
 \ea
 ${\vec{\bf{n}}}\,$ being the outward normal and $\,dS\,$ being an element of the surface area of the primary. This expression for the
 power (pioneered, probably, by Goldreich 1963) enables one to calculate the work through radial displacements only, in neglect of
 horizontal motion.

 {}

\end{document}